\documentclass[prb,twocolumn,showpacs,amsmath,amssymb,superscriptaddress]{revtex4}% Physical Review B
\newcommand\beq{\begin{equation}}
\newcommand\eeq{\end{equation}}

\usepackage{dcolumn}% Align table columns on decimal point
\usepackage{bm}% bold math

\usepackage{subfigure}
\usepackage{amsmath}
\usepackage{amsfonts}   % if you want the fonts
\usepackage{amssymb}
\usepackage[final]{graphicx}

\begin{document}

%\preprint{APS/123-QED}

\title{Directive Emission from Defect-Free Dodecagonal Photonic Quasicrystals:
\\A Leaky-Wave Characterization}

\author{Alessandro Micco}
\author{Vincenzo Galdi}\email{vgaldi@unisannio.it}
\affiliation{Waves Group, Department of Engineering, University of Sannio, I-82100 Benevento, Italy}
\author{Filippo Capolino}
\affiliation{Department of Electrical Engineering and Computer Science, University of California, Irvine, CA 92697, USA}
\author{Alessandro Della Villa}
\affiliation{Department of Information Engineering, University of Siena, I-53100 Siena, Italy}
\author{Vincenzo Pierro}
\affiliation{Waves Group, Department of Engineering, University of Sannio, I-82100 Benevento, Italy}

\author{Stefan Enoch}
\author{G\'erard Tayeb}
\affiliation{Institut Fresnel, CNRS, Aix-Marseille Universit\'e, 13397 Marseille cedex 20, France}

\date{\today}% It is always \today, today,

%%%%%%%%%%%%%%%%%%%%%%%%%%%%%%%%%%%%%%%%%%%%%%%%%%%%
\begin{abstract}
%%%%%%%%%%%%%%%%%%%%%%%%%%%%%%%%%%%%%%%%%%%%%%%%%%%%
In this paper, we study the radiation from
embedded sources in two-dimensional finite-size 
``photonic-quasicrystal'' (PQC) slabs made of dielectric rods arranged according to  
a 12-fold symmetric aperiodic tiling. The results from our investigation, based on rigorous full-wave
simulations, show the possibility of achieving broadside radiation at multiple frequencies, with high-directivity (e.g., 15 dB) and low-sidelobes (e.g., -12 dB).  
We also show that leaky waves are supported by a PQC slab, and that the beamwidth is directly proportional to the leaky-wave attenuation constant, which provides a physically-incisive interpretation of the observed radiation characteristics.
\end{abstract}

\maketitle

%%%%%%%%%%%%%%%%%%%%%%%%%%%%%%%%%%%%%%%%%%%%%%%%%%%%%%%%%%%
\section{Introduction and Background}
%%%%%%%%%%%%%%%%%%%%%%%%%%%%%%%%%%%%%%%%%%%%%%%%%%%%%%%%%%%
The study of the physical properties and applications of {\em aperiodically-ordered} structures, related to the concept of ``quasicrystals'' in solid-state physics \cite{Senechal1}, is eliciting a growing attention in many fields of science and technology \cite{Macia}. In electromagnetics (EM) and optical engineering, attention has been focused on {\em photonic quasicrystals} (PQCs), mainly inspired by the theory of ``aperiodic tilings'' \cite{Senechal1}, i.e., arrangements of polygonal shapes devoid of any translational symmetry and capable of covering a plane without overlaps or gaps. Although aperiodic, these geometries can still exhibit long-range order and (high-order, ``noncrystallographic'') rotational symmetry in a local/statistical sense \cite{Senechal1}.
A large body of PQC studies, in applications ranging from lasers to superfocusing (see Ref. \onlinecite{Steurer} for a recent review), have demonstrated the possibility of obtaining similar effects as those exhibited by periodic photonic crystals (PCs), with intriguing potentials (e.g., bandgap control, lower/multiple operational frequencies, higher isotropy, richer defect-state spectrum, etc.) stemming from the additional degrees of freedom endowed by aperiodic order.

One application that still remains largely unexplored concerns the possibility of achieving directive emission from embedded sources in PQCs. In periodic PCs, such possibility has been demonstrated in several investigations, and its theoretical aspects are well-understood. 
In particular, directive emission has been achieved via the use of microcavities \cite{Tayeb} (and, possibly, slight local changes in the lattice period), or planar defects \cite{Thevenot,Cheype,Ozbay1}, as well as in  defect-free configurations \cite{Ozbay,Enoch3} by suitably designing the spatial dispersion properties. Periodic structures interpreted as artificial materials \cite{Brown}
characterized by an effective relative permittivity lower than one have also been proposed as effective devices to increase the directivity 
\cite{Gupta1,Bahl,Enoch1}. Building up on the seminal work in Ref.
\onlinecite{Tamir}, the underlying phenomenology was also explained in terms of
a {\em leaky-wave} (LW) model in Refs. \onlinecite{Gupta1,Bahl} (see also Sect. \ref{Sec:LW}). In a
similar framework, a detailed analysis and parameterization
capable of predicting quantitatively the directivity enhancement
was derived in Ref. \onlinecite{Lovat} for a periodic wire medium
slab, in terms of very fast and slowly-attenuating LWs.

For PQCs, the lack of spatial periodicity (and related Bloch-type analytic tools) significantly complicates the modeling and understanding of the involved phenomenologies (see, e.g., Ref. \onlinecite{DellaVilla2005,DellaVilla2006}).
In Ref. \onlinecite{DellaVilla2006a}, we presented a preliminary study of the radiation from an electric line-source embedded in a two-dimensional (2-D) finite-size dielectric PQC slab with Penrose-type (5-fold symmetric, quasiperiodic \cite{Senechal1}) lattice.
Results showed the possibility of achieving moderate directivity at three frequencies. Interestingly, the lowest operational frequency was found to be moderately smaller than the corresponding value observed for a periodic PC slab with comparable size and number of rods, but the side-lobe level (SLL) was still rather high. It is worth pointing out that the physical interpretation of such results was not fully clear, although the possible applicability of LW-based models as in Ref. \onlinecite{Lovat} was envisaged.

In this paper, building up on the preliminary study in Ref. \onlinecite{DellaVilla2006a}, we further elaborate upon this subject, by exploring
some of the open issues related to: {\em i)} the possible extension to other tiling geometries, {\em ii)} the physical interpretation, and {\em iii)} the parametric optimization (in particular, the SLL reduction).
In this framework, we consider a different tiling geometry with
12-fold symmetry \cite{Oxborrow,Zoorob1}, recently explored in
connection with superfocusing
\cite{Feng1,DiGennaro1,DiGennaro}. Via a parametric study based on full-wave
numerical simulations \cite{Tayeb}, we find several instances of
directive emission with low SLL, and address their physically-incisive 
interpretation in terms of a LW model similar to that proposed in Ref. \onlinecite{Lovat}, highlighting similarities and differences with the periodic PC case.

Accordingly, the rest of the paper is organized as follows. In Sect. \ref{Sec:Geom}, we outline the problem geometry and the observables utilized throughout. In Sect. \ref{Sec:Results}, we illustrate some representative results, address their physical interpretation, and summarize the results from our parametric studies. Finally, in Sect. \ref{Sec:Conclusions}, we provide some brief concluding remarks and hints for future research.

%############################################################
%                Figure1
%
\begin{figure}
\begin{center}
\includegraphics[width=8cm]{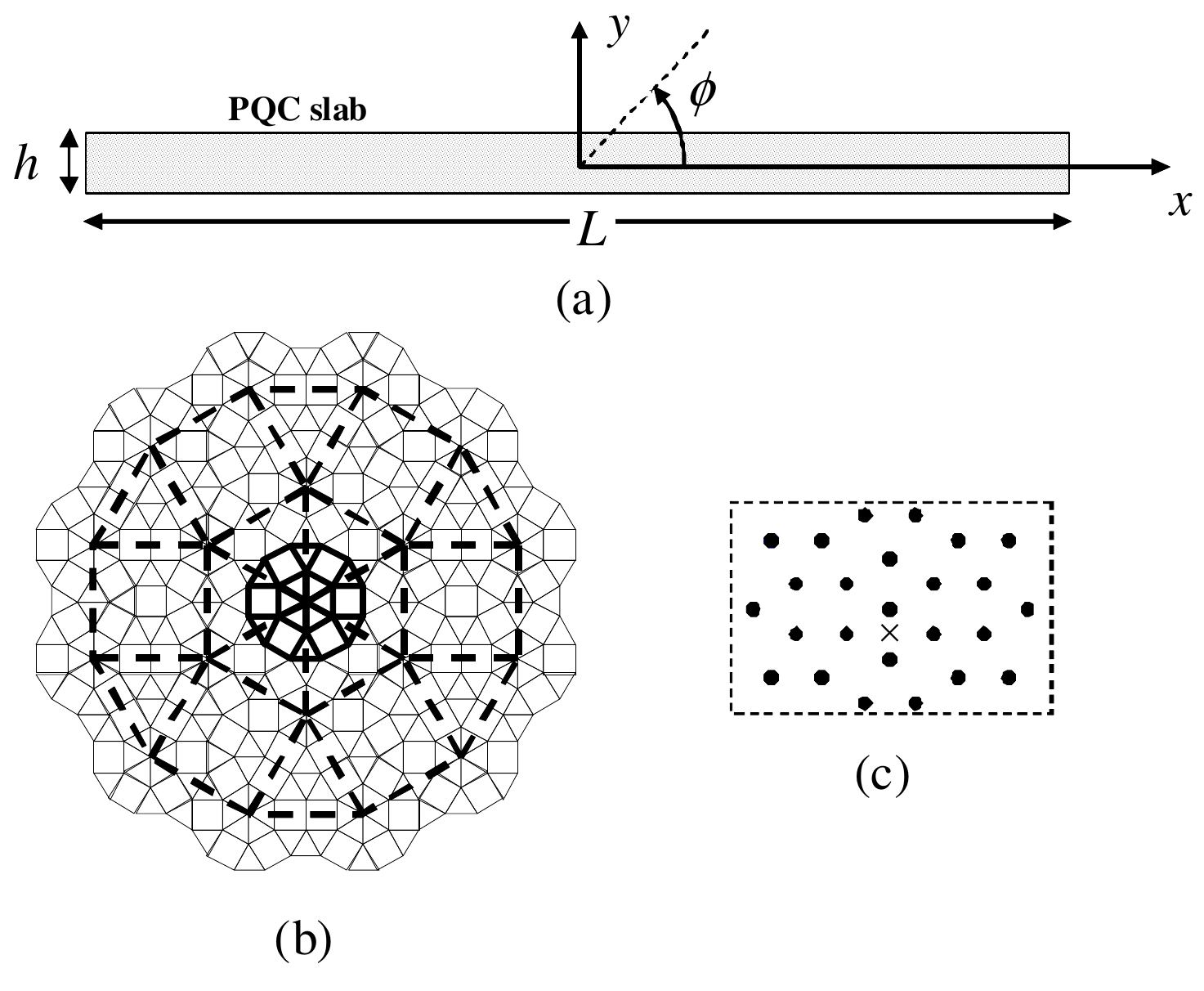}
\end{center}
\caption{(a): Problem geometry. (b): Stampfli inflation rule. (c): Zoom of the central slab area (bullets and cross mark the positions of dielectric rods and line source, respectively). Details are explained in the text.}
\label{Figure1}
\end{figure}
%#############################################################

%%%%%%%%%%%%%%%%%%%%%%%%%%%%%%%%%%%%%%%%%%%%%%%%%%%%%%%%%%%%%%
\section{Geometry and Observables}
%%%%%%%%%%%%%%%%%%%%%%%%%%%%%%%%%%%%%%%%%%%%%%%%%%%%%%%%%%%%%%
\label{Sec:Geom}
As in Ref. \onlinecite{DellaVilla2006a}, we consider a 2-D ($z$-invariant) geometry (Fig. \ref{Figure1}(a)) involving a PQC slab based on a
dodecagonal aperiodic tiling (in the $x-y$ plane), characterized by local and statistical 12-fold rotational symmetry, and generated via the so-called Stampfli inflation rules \cite{Oxborrow} illustrated in Fig. \ref{Figure1}(b).
The construction, to be iterated up to the desired tiling extension, starts from a dodecagon made of square and equilater-triangle tiles (thick solid line), which is first scaled-up by an inflation factor $\sqrt{3}+2$ (thick dashed line), and subsequently substituted at each vertex of the inflated version (see Refs. \onlinecite{Zoorob1,Feng1,DiGennaro1,DiGennaro,Jin2,Chan2,Gauthier} for similar and different 12-fold symmetric PQCs). The PQC slab is then generated by cutting a rectangular portion of size $L\times h$ from a suitably large tiling, and placing circular dielectric rods (in vacuum) at the tile vertices; no defects are assumed. The rods have relative permittivity $\epsilon_r=12$ and radius $r=0.138a$, with $a$ being the lattice constant (tile sidelength). A time-harmonic ($\exp(-i\omega t)$) electric line-source is placed in vacuum nearby the slab center (see the zoom in Fig. \ref{Figure1}(c)).

As meaningful observables, we consider the normalized ``local density of states'' (LDOS) \cite{Asatryan}
\beq
\rho({\bf r}_s,\omega) = 4
\textrm{Im} \{ G({\bf r}_s,{\bf r}_s;\omega)\},
\label{eq:LDOS}
\eeq
as well as the directivity
\beq
D(\phi,\omega)=\frac{2\pi|E^{ff}_s(\phi,\omega)|^2}{\int_{0}^{2\pi}|E^{ff}_s(\phi,\omega)|^2 d\phi},
\label{eq:D}
\eeq
the normalized intensity
\beq
{\bar S}(\phi,\omega)=\frac{|E^{ff}_s(\phi,\omega)|^2}{|E^{ff}_{s0}(\phi,\omega)|^2},
\label{eq:S}
\eeq
and the SLL
\beq
\Lambda(\omega)=\frac{\max_{\phi \notin {\cal B}}|E^{ff}_s(\phi,\omega)|^2}
{\max_{\phi\in[0,180^o]}|E^{ff}_s(\phi,\omega)|^2},
\label{eq:SLL}
\eeq
with $G$ denoting the PQC Green's function relating the current to the magnetic vector potential, $E^{ff}_s$ and $E^{ff}_{s0}$ the far-fields radiated by the line-source in the presence and absence, respectively, of the PQC, and ${\cal B}$ the main lobe angular region.
The above observables, equal to one in the absence of the PQC, are computed via a well-established full-wave numerical technique based on a multipolar Fourier-Bessel expansion \cite{Tayeb}. In order to facilitate comparisons with previously published results, we also consider the 3dB beamwidth $\Delta\phi_{3dB}$, defined as the angular region where the radiation is higher than -3dB of the maximum, i.e. (for a symmetric radiation pattern with main lobe at $\phi=90^o$),
\beq
|E_s^{ff}(\pi/2+\Delta\phi_{3dB}/2,\omega)|=|E_s^{ff}(\pi/2,\omega)|/\sqrt{2}.
\eeq
In the 2-D scalar scenario of interest here, the LDOS admits an intuitive interpretation as the normalized {\em total} power (per unit-length along the $z$ direction)
radiated by the line-source \cite{DellaVilla2006a}, and represents a meaningful bandgap indicator, whereas the directivity and the normalized intensity are indicative of the angular confinement and power density enhancement (as compared to vacuum) of the radiated far-field.
Attention is focused on achieving {\em broadside} ($\phi=90^o$) directivity, although the structure also radiates into the $y<0$ half-space.

%############################################################
%                Figure2
%
\begin{figure}
\begin{center}
\includegraphics[width=8cm]{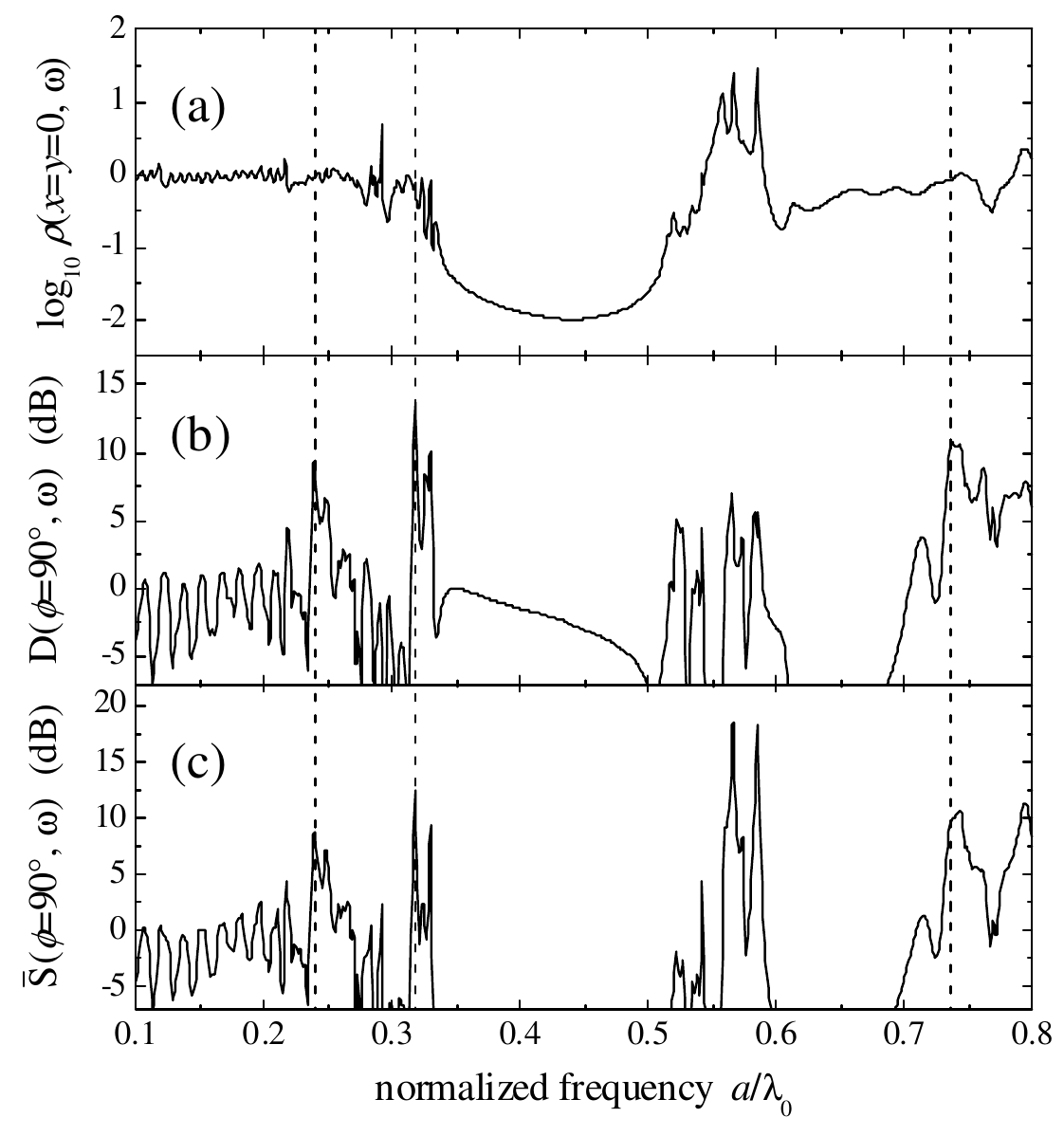}
\end{center}
\caption{Normalized LDOS (a), broadside directivity (b), and broadside normalized intensity (c), as a function of the normalized frequency $a/\lambda_0$, for a dodecagonal PQC slab ($L=97.8a$, $h=4a$, 435 rods, cf. Fig. \ref{Figure1}c), with source at $x=0$, $y=-0.5a$.}
\label{Figure2}
\end{figure}
%#############################################################

%############################################################
%                Figure3
%
\begin{figure}
\begin{center}
\includegraphics[width=7cm]{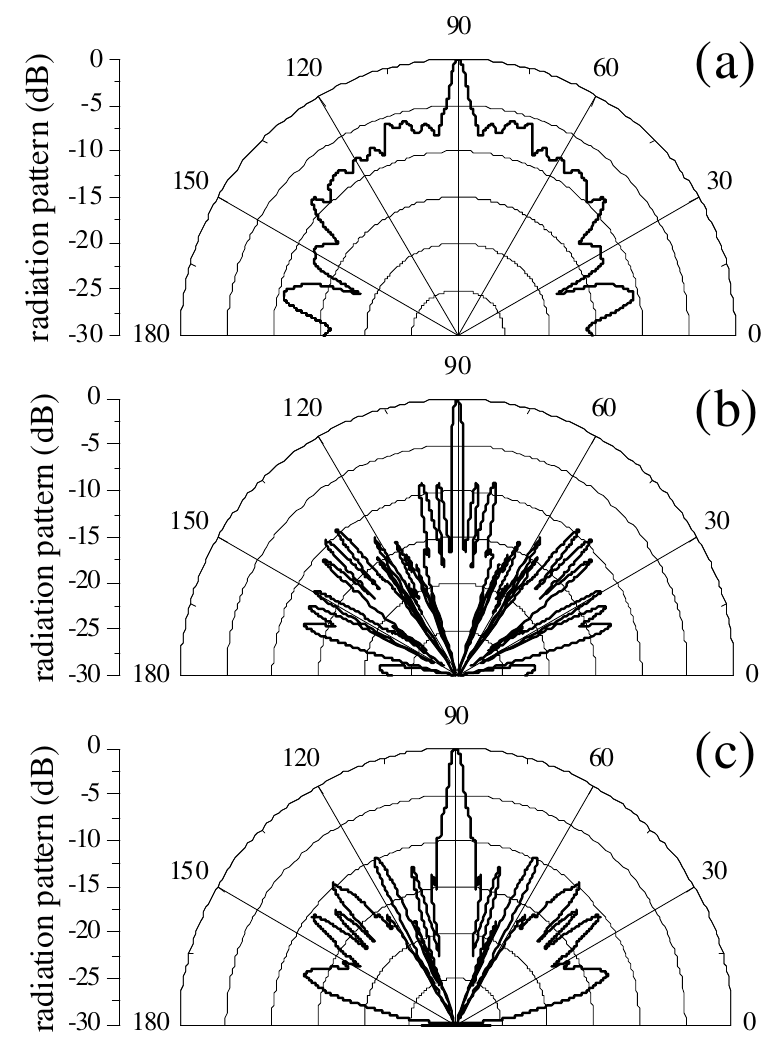}
\end{center}
\caption{Radiation patterns (in the upper half-space) for a configuration with the same parameters as in Fig. \ref{Figure2}, at three normalized frequencies (marked as vertical dashed lines in Fig. \ref{Figure2}). (a), (b), (c): $a/\lambda_0=0.239, 0.317$, and $0.736$, respectively.}
\label{Figure3}
\end{figure}
%#############################################################

%%%%%%%%%%%%%%%%%%%%%%%%%%%%%%%%%%%%%%%%%%%%%%%%%%%%%%%%%%%%%%
\section{Directive Emission from a Dodecagonal PQC Slab}
%%%%%%%%%%%%%%%%%%%%%%%%%%%%%%%%%%%%%%%%%%%%%%%%%%%%%%%%%%%%%%
\label{Sec:Results}

%-------------------------------------------------------------
\subsection{Representative Results}
%-------------------------------------------------------------
Figures \ref{Figure2}(a)--\ref{Figure2}(c) show the observables in (\ref{eq:LDOS})--(\ref{eq:S}) as a function of the normalized frequency $a/\lambda_0$ (with $\lambda_0$ denoting the vacuum wavelength). From the LDOS response (Fig. \ref{Figure2}(a)), one can identify a main bandgap (represented by low values of $\rho$) and several minor dips. Potentially interesting configurations are characterized by high-amplitude peak coincidences in the broadside directivity (Fig. \ref{Figure2}(b)) and normalized intensity (Fig. \ref{Figure2}(c)). These coincidences correspond to situations where the peaks in the radiated power (Fig. \ref{Figure2}(c)) are likely accompanied by radiation patterns with narrow beams, and thus high directivity, in the $\phi=90^o$ direction.
The vertical dashed lines mark three such representative configurations, whose radiation patterns (normalized $|E^{ff}_s(\phi,\omega)|^2$ vs. $\phi$) are displayed in Figs. \ref{Figure3}(a)--\ref{Figure3}(c). In all three cases, one should observe the narrow peak at broadside accompanied by sidelobes of lower level. It should be emphasized that, as clearly visible from the sharp peaks in Fig. \ref{Figure2}, the above configurations are intrinsically {\em narrow-band}.
At the lowest frequency ($a/\lambda_0=0.239$), the directivity is about 9dB (with 3dB beamwidth $\Delta\phi_{3dB}=4^o$), and the SLL (Fig. \ref{Figure3}(a)) is moderately high ($\Lambda\approx-6$ dB), comparable with the values observed in Ref. \onlinecite{DellaVilla2006a} for a Penrose-type PQC. However, at the higher frequencies $a/\lambda_0=0.317, 0.736$, the radiation patterns look {\em considerably cleaner}, with directivities of 13.7dB ($\Delta\phi_{3dB}=2.35^o$) and 10.4dB ($\Delta\phi_{3dB}=3.44^o$), respectively, and the SLL reaches values $\sim-10$dB that may be acceptable in many practical applications. 
As compared with the Penrose tiling geometry in Ref. \onlinecite{DellaVilla2006a}, the dodecagonal PQC here turns out to provide comparable operating frequencies and lower SLL.   

It is worth noticing that direct comparisons of above results with the periodic PC cases in Refs. \onlinecite{Tayeb,Thevenot,Cheype,Ozbay1,Ozbay,Enoch3} should be addressed cautiously, in view of the differences in the underlying radiation mechanisms and related parametric ranges. In this connection, we highlight that the configurations in Refs. \onlinecite{Tayeb,Thevenot,Cheype,Ozbay1,Ozbay,Enoch3} are based on either defected or defect-free {\em thick} structures, which essentially exploit the spatial filtering capabilities of PCs around band-edge frequencies in order to achieve directive emission. Conversely, our configuration is based on a defect-free thin-slab geometry, and, as one can observe from Fig. \ref{Figure2}, the operational frequencies are not necessarily close to the band edges. 
As a reference for {\em qualitative} comparisons, we observe from Ref. \onlinecite{Ozbay} that the radiation pattern produced by a line-source embedded in a defect-free periodic PC thick-slab made of $24\times16$ alumina ($\epsilon_r=9.61$) rods of radius $r=0.136a$ (in the experiment, it was chosen $a=1.1$cm and $r=1.5$mm), at a frequency of 13.21 GHz (i.e., $a/\lambda_0=0.484$), would exhibit a 3dB beamwidth $\Delta\phi_{3dB}=8^o$, and SLL values $\sim -20$dB. Therefore, from a qualitative viewpoint (and bearing in mind the aforementioned parametric differences), the performance exhibited by our proposed configuration (at the higher frequencies) turns out to be comparable, with moderately smaller beamwidth (higher directivity) and higher SLL.

We also would like to point out, that in general, the directivity of thick structures is governed by geometrical-optics ultra-refraction phenomena, as explained in Ref. \onlinecite{Enoch1} for a metamaterial with a rectangular lattice and in Ref. \onlinecite{{Lovat_3}} for its equivalent homogenized low-permittivity material. On the other hand, thin slabs are able to provide higher directivitiy because the physical mechanism is slightly different, as explained in Ref. \onlinecite{{Lovat_3}}. There, it is shown that, because of the small thickness, a (single) dominant LW can be excited. In Ref. \onlinecite{{Lovat_3}} and in Sect. III-D of Ref. \onlinecite{{Lovat_2}}, the two mechanisms (geometrical-optics-based for thick structures, and LW-based for thin structures) have been quantitatively compared, showing that the latter is capable of yielding a higher directivity.

%############################################################
%                Figure4
%
\begin{figure}
\begin{center}
\includegraphics[width=8.5cm]{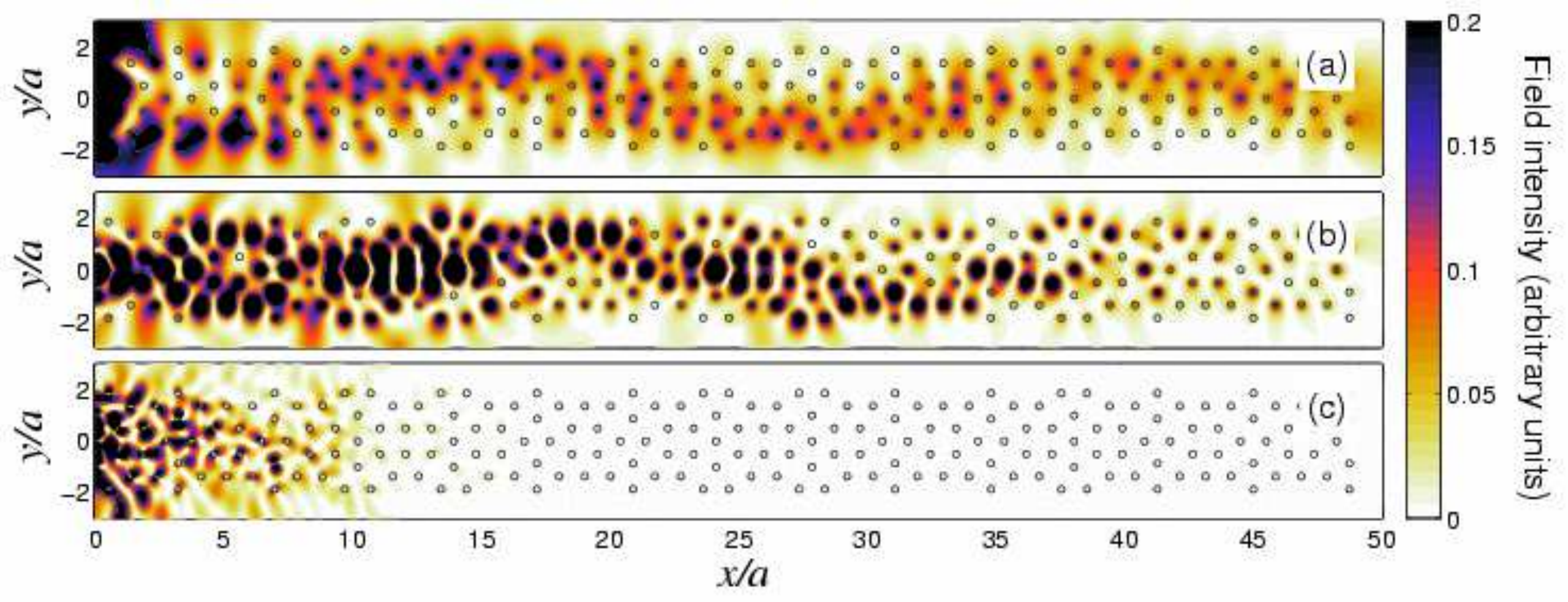}
\end{center}
\caption{(Color online) Internal field intensity maps pertaining to the radiation patterns in Fig. \ref{Figure3} (due to symmetry, only the $x>0$ slab-half is shown). (a), (b), (c): $a/\lambda_0=0.239, 0.317$, and $0.736$, respectively.}
\label{Figure4}
\end{figure}
%#############################################################

%############################################################
%                Figure5
%
\begin{figure}
\begin{center}
\includegraphics[width=8cm]{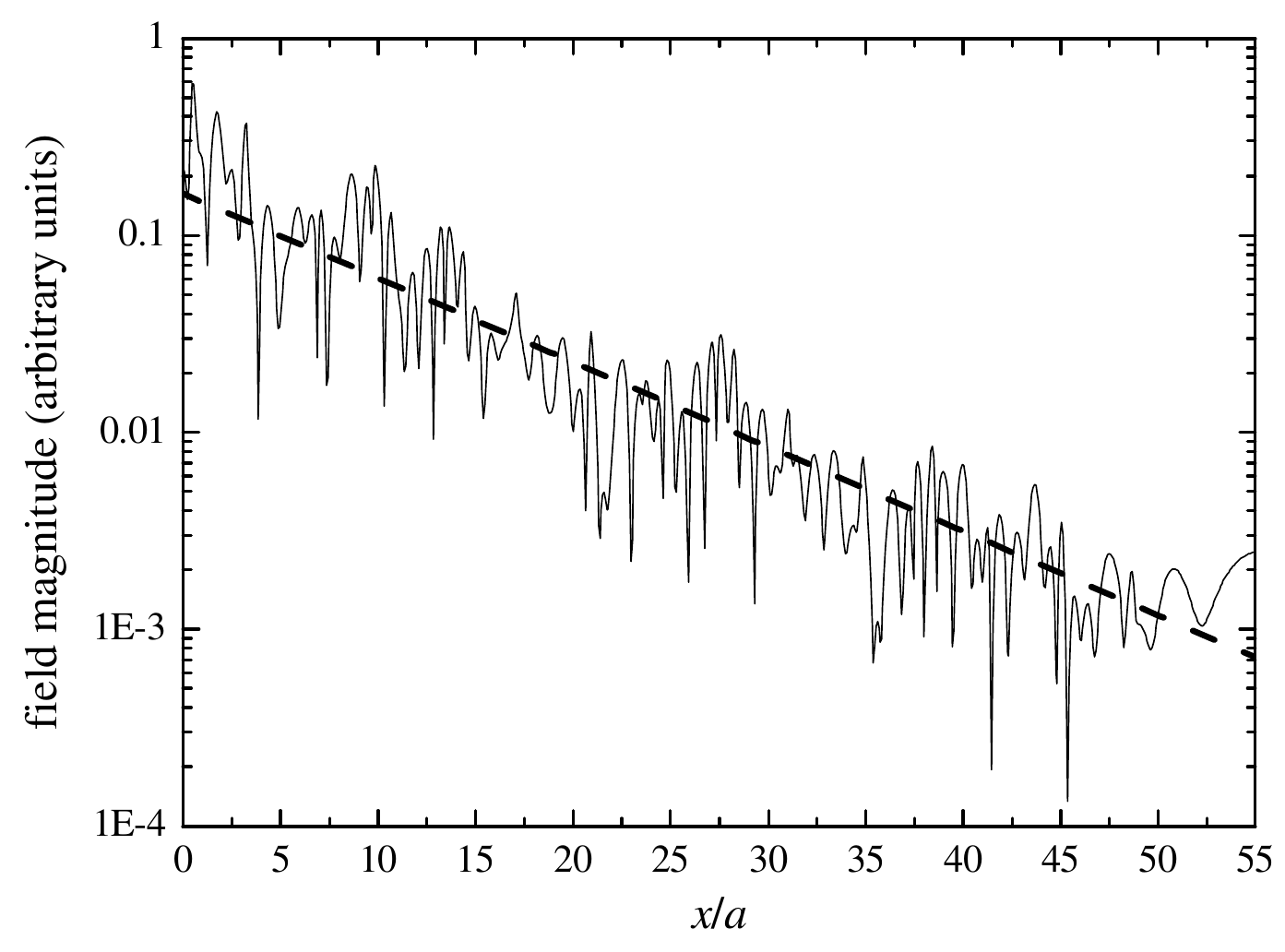}
\end{center}
\caption{Field-magnitude cut at $y=2.03a$ for the configuration in Fig. \ref{Figure4}(c) (corresponding to the radiation pattern in Fig. \ref{Figure3}(c)). Note the semi-log scale and the linear  fit (shown
dashed), which highlight the exponential field decay along $x$.}
\label{Figure5}
\end{figure}
%#############################################################

%-------------------------------------------------------------
\subsection{Physical Interpretation}
%-------------------------------------------------------------
\label{Sec:LW}
In this Section, we show that the radiation mechanism underlying the observed results is related to the excitation of a LW that travels along the PQC slab. A LW is a guided wave that radiates continuously across a guiding structure while propagating
along it; the power leakage results in an exponential attenuation of the wave as
it propagates along the structure, even in the absence of losses. 

The radiation mechanism of our PQC thin slab turns out to be somehow similar to those illustrated in Refs. \onlinecite{Gupta1,Bahl,Enoch1}, which can be effectively parameterized via a LW model (see Refs. \onlinecite{Tamir,Lovat,Capolino,Jackson_1,Lovat_2} for more details).
For slabs made of periodic structures, it has been observed \cite{Lovat} that the field excited by a source and evaluated at a certain distance from the vacuum-slab interface is mainly represented by a LW mode contribution. Such distance from the interface is required in order to observe a homogeneous field distribution that is not strictly related to the local structure details described by evanescent waves (associated with high-order Bloch harmonics) that decay away from the interface. Along the interface, at a fixed $y$, a general description of the field in terms of algebraically-decaying spatial waves and guided (bounded and leaky) modes is given in Ref. \onlinecite{Tamir}, Sect. III of Ref. \onlinecite{Lovat}, and Sect. V of  Ref. \onlinecite{Capolino}. In certain structures (see, e.g., those shown in Ref. \onlinecite{Capolino}), the spatial wave is the dominant contribution. However, in other cases (which can be properly designed \cite{Lovat}), the spatial wave is comparatively weakly excited, and the near-field $E_s$ at an equivalent aperture close to the vacuum-PC interface is therefore dominated by a LW mode with complex propagation constant $k_{LW}=\beta+i\alpha$, i.e., 
\beq
E_s(x,y)\sim E_0(y) \exp\left(i \beta |x|\right) \exp\left(-
\alpha |x|\right),
\label{eq:LW} 
\eeq 
where $\beta$ is the phase propagation constant, and the attenuation constant $\alpha$ describes the exponential decay along $x$. If the attenuation constant $\alpha$ is much smaller than the free space wavenumber $k_0=2\pi/\lambda_0$,  the field at the vacuum-PC interface is sustained for a long distance, extending for several wavelengths. If also the phase propagation constant $\beta$ is much smaller than the free space wavenumber (i.e., $\beta\ll k_0$), we derive that the field at the interface exhibits an essentially equi-phase profile, which is an important aspect for high-directivity radiators. 
It is important to note that close to the vacuum-PC interface (at distances smaller or comparable with the lattice constant) the field oscillates around the average value represented by (\ref{eq:LW}) because of the effects produced by the nearby scatterers (for periodic structures, these oscillations are represented by high-order Bloch harmonics).

As anticipated, another important outcome of this study is the evidence that LWs can be supported by a PQC slab as well. In order to gain such evidence, we show that the field at the vacuum-PQC interface can be described by the simplified model in (\ref{eq:LW}).
As a first example, Fig. \ref{Figure4} shows the intensity field-maps inside the PQC slab at the
three representative frequencies, from which the presence of a modal structure is clearly visible.
%############################################################
%                Figure6
%
\begin{figure}
\begin{center}
\includegraphics[width=7cm]{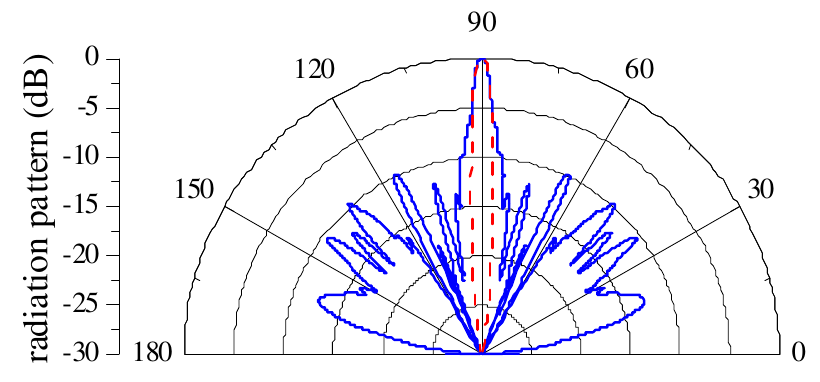}
\end{center}
\caption{(Color online) Comparison between the actual radiation pattern at normalized frequency $a/\lambda_0=0.736$ (blue solid curve, cf. Fig. \ref{Figure3}(c)) and LW-model prediction for $\beta=\alpha=0.099/a$ (red dashed curve, cf. (\ref{eq:LW_RP1})).}
\label{Figure6}
\end{figure}
%#############################################################

The LW propagation constant $k_{LW}$ is very useful to parameterize the characteristics of the radiated field, which can be readily computed via Fourier transform of the LW aperture field in (\ref{eq:LW}). We can therefore write the LW radiation pattern as \cite{Lovat,Jackson_1}
\begin{eqnarray}
&&\left|E_{LW}^{ff}(\phi,\omega)\right|^2\sim A \sin^2\phi\nonumber\\
&&~~~~~\times\left\{ \frac{\beta^2-\alpha^2+4\alpha\beta}{\left[k_0^2\cos^2\phi-\left(\beta^2-\alpha^2\right)^2\right]^2+\left(2\alpha\beta\right)^2}\right\},
\label{eq:LW_RP}
\end{eqnarray}
where $A$ is a normalization constant. From (\ref{eq:LW_RP}), it can been shown that a radiation peak at broadside is obtained
 when \cite{Lovat,Lovat_2}  $\beta \ll k_0$ and $\beta \le
\alpha$. Moreover, the highest directivity at broadside is found
when the optimum condition $\beta \approx \alpha$ is
satisfied \cite{Lovat,Lovat_2}, in which case the radiation pattern becomes
\beq
\left|E_{LW}^{ff}(\phi,\omega)\right|^2\sim A \sin^2\phi\left(\frac{4\alpha^2}{k_0^4\cos^4\phi+4\alpha^4}\right).
\label{eq:LW_RP1}
\eeq
From this parameterized expression of the radiation pattern, we can derive the 3dB beamwidth,  
\beq
\Delta\phi_{3dB}\approx2\arcsin\left(\frac{\sqrt{2}\alpha}{k_0}\right)
\approx \frac{2\sqrt{2}\alpha}{k_0}.
\label{eq:phi3dB}
\eeq 
At variance of periodic PCs and artificial materials,
in our aperiodic geometry, the LW propagation constant $k_{LW}$ cannot be
systematically computed from a dispersion equation. Nevertheless,
we can show that the observed radiation features are consistent with the predictions of a simplified LW model as in
(\ref{eq:LW})--(\ref{eq:phi3dB}). To this aim, for the operational frequency
$a/\lambda_0=0.736$, Fig. \ref{Figure5} shows a cut of the field
magnitude at an aperture $y=2.03 a$ close to the vacuum-PQC
interface. Thanks to the semi-log scale
and the dashed linear fit in the plot, an exponential decay is clearly observable, with
irregular oscillations around the linear fit ascribable to the inherent local
structure of the PQC slab interface. The LW attenuation
constant $\alpha$ can readily be estimated from the slope of the
log-scale linear fit, yielding $\alpha\approx 0.099/a$, which
 is consistent ($\alpha \ll k_0$) with the
observed high directivity \cite{Lovat}. Due to the aforementioned lack of a dispersion equation, we are not able to compute $\beta$. Nevertheless, the observation from Fig. \ref{Figure2}(c) that at the chosen frequency $a/\lambda_0=0.736$ the directivity is actually {\em maximum} suggests
that the condition $\beta\approx\alpha$ is satisfied (see the discussion after (\ref{eq:LW_RP})).
This is confirmed by the comparison, shown in Fig. \ref{Figure6}, between the actual radiation pattern (cf. Fig. \ref{Figure3}(c)) and the LW prediction from (\ref{eq:LW_RP1}), which evidences a good agreement in the angular region near the maximum; the two curves are almost coincident over a 5 dB dynamic range. From a more quantitative viewpoint, Eq. (\ref{eq:phi3dB}) estimates the corresponding beamwidth  $\Delta\phi_{3dB}=3.46^o$, in excellent
agreement with the actual value of $3.44^o$. 
We can thus conclude that the observed radiation properties are indeed consistent with the predictions of the simplified LW model in (\ref{eq:LW})--(\ref{eq:phi3dB}), and hence a LW with the appropriate characteristics (namely, with amplitude and phase profiles varying on a much larger length scale than the vacuum wavelength) is mainly responsible for shaping the main beam of the pattern.

Another interesting feature offered by the LW parameterization is the information about the transverse size (along the $x$ direction) that is necessary to produce a required beam radiation. Once the beamwidth $\Delta\phi_{3dB}$ is specified, the required $\alpha=\beta$ attenuation constant is then found via (\ref{eq:phi3dB}). From (\ref{eq:LW}), one can then determine how large the structure needs to be for having a negligible field at its edges, so that truncation effects would not produce pattern distortion in the main lobe region.

%############################################################
%                Figure7
%
\begin{figure}
\begin{center}
\includegraphics[width=7cm]{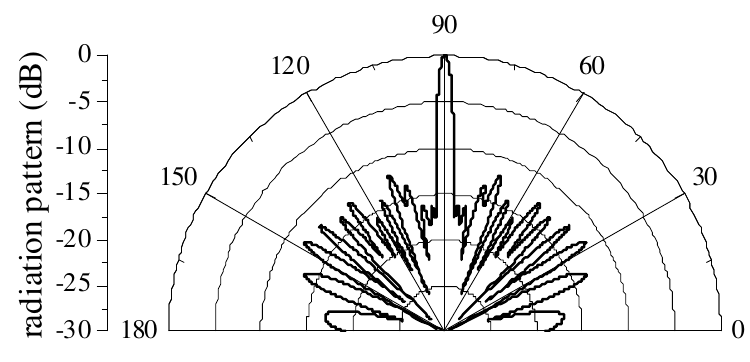}
\end{center}
\caption{As in Fig. \ref{Figure3}(a), but with source at $x=0$, $y=1.55a$.}
\label{Figure7}
\end{figure}
%#############################################################

%-------------------------------------------------------------
\subsection{Summary of Parametric Studies}
%-------------------------------------------------------------
From the above physical interpretation, one gathers that directive low-SLL radiation can be achieved by exciting a proper LW mode in the PQC slab.
In order to gain further insight into the role of the various parameters involved in the excitation process, we carried out a comprehensive set of parametric studies, whose main results can be summarized as follows. 

\begin{itemize}
\item{The symmetric-cut configuration in Fig. \ref{Figure1} turns out to be the best suited for exciting a LW radiating at broadside. Different cuts from the original dodecagonal tiling, where either the $x-$ or $y-$symmetry is broken, can be utilized for pointing the main beam in different directions (as already observed in Ref. \onlinecite{DellaVilla2006a}).}

\item{Increasing the slab thickness, we also observed variations in the radiation characteristics, though without the clearly-predictable behavior observable in the periodic PC case \cite{Gupta1,Bahl,Enoch1,Lovat}. In the periodic PC case, it is known that increasing too much the slab thickness, other LW modes may appear and deteriorate the radiation pattern. In this connection, it should also be noticed that, unlike in periodic PCs, thickness changes in PQC slabs intrinsically involve {\em termination} changes, and so the two effects (thickness and termination) are not easily discernible.}

\item{The rods radius value ($r=0.138a$) utilized throughout was arrived at via parameter scanning. Slight departures from this value (e.g., $r=0.15a$) were observed to yield moderate deterioration in the SLL, up to the complete disappearance of the directive response at the higher frequencies.}

\item{The source position was found to be a rather effective parameter for adjusting the LW excitation, so as to optimize the radiation pattern. As an example, the radiation pattern in Fig. \ref{Figure3}(a) was optimized by moving the source vertically (so as to preserve the $x$-symmetry and, hence, the broadside radiation) via parameter scanning. Figure \ref{Figure7} shows the optimized radiation pattern, which represents the best tradeoff response ($D=15.5$dB, ${\bar S}=14.8$dB, and $\Lambda=-12$dB) observed for a dodecagonal PQC in our study.}

\end{itemize}

%%%%%%%%%%%%%%%%%%%%%%%%%%%%%%%%%%%%%%%%%%%%%%%%%%%%%%%%%%%%%%
\section{Conclusions}
%%%%%%%%%%%%%%%%%%%%%%%%%%%%%%%%%%%%%%%%%%%%%%%%%%%%%%%%%%%%%%
\label{Sec:Conclusions}
In this paper, we have presented a study of the radiation properties of line-sources embedded in 2-D finite-size defect-free dielectric dodecagonal PQC slabs. Via a comprehensive full-wave numerical study, we have identified certain parametric configurations which give rise to directive low-SLL broadside radiation. These results extend to a different PQC geometry our previous observations \cite{DellaVilla2006a}, confirming the possibility of obtaining directive low-SLL radiation at multiple frequencies.
Moreover, we have shown that also for the case of {\em aperiodically-ordered} PQCs the observed radiation characteristics are consistent with the predictions of a LW-based simplified model, and have explored the role of the key geometrical parameters.

The above results, and their LW-based physical interpretation, pave the way for a systematic parameterization and design of directive radiators based on PQC slabs. In this framework, current and future investigations are aimed at a deeper understanding of the modal structure supported by PQC slabs, and the refinement of the LW-based modeling, so as to come up with more quantitative design rules. Also of interest is the experimental verification of the phenomenon, as well as further studies of different configurations (e.g., air-hole-type) and PQC geometries.

%%%%%%%%%%%%%%%%%%%%%%% References %%%%%%%%%%%%%%%%%%%%%%%%%

%%%%%%%%%%%%%%%%%%%%%%%%%%%%%%%%%%%%%%%%%%%%%%%%%%%%%%%%%%%%%%
\section*{Acknowledgments}
%%%%%%%%%%%%%%%%%%%%%%%%%%%%%%%%%%%%%%%%%%%%%%%%%%%%%%%%%%%%%%
This work was supported in part by the Italian Ministry of Education and
Scientific Research (MIUR) under the PRIN-2006 grant on ``Study and
realization of metamaterials for electronics and TLC applications,''
and in part by the Campania Regional Government via a 2005 grant (L.R. N. 5 -- 28.03.2002) on ``Electromagnetic-bandgap quasicrystals: Study, characterization, and applications in the microwave region.''

The authors wish to thank two anonymous reviewers for useful comments and suggestions.

\newpage

\end{document}